\documentclass[12pt,a4paper]{article}
\usepackage[utf8]{inputenc}
\usepackage{graphicx}
\usepackage{xcolor}
\usepackage{tikz}
\usepackage{tikz-3dplot}
\usetikzlibrary{decorations.pathmorphing,arrows.meta}
\usetikzlibrary{calc}
\usepackage{amsmath}
\usepackage{amssymb}
\usepackage{hyperref}
\usepackage{geometry}
\usepackage{caption}
\usepackage{subcaption}

\usepackage{authblk}

\tdplotsetmaincoords{70}{110}

\title{Quantifying Quantum Correlations in Annihilation Photon Pairs under Compton Scattering}
\author[1]{Z. AskariPour Ravari\footnote{Corresponding author: zakiehaskary@gmail.com}}
\author[2]{Z. Riazi}
\affil[1]{Department of Physics, Islamic Azad University, North Tehran Branch, Tehran, Islamic Republic of Iran}
\affil[2]{Physics and Accelerator Research School, Nuclear Science and Technology Research Institute, Tehran, Iran}
\date{\today}

\begin{document}
\maketitle

\begin{abstract}

We present a theoretical study of the evolution of polarization entanglement and quantum coherence in 511 keV photon pairs produced by para-positronium decay during successive Compton scattering events. We start with a maximally entangled Bell state and employ the generalized Stokes-Mueller formalism to derive the two-photon density matrix following single-, double-, and triple-Compton scattering, explicitly considering both polar and azimuthal scattering geometries. Using this framework, we quantify the degradation of quantum correlations through concurrence (as a measure of entanglement) and the $l_1$-norm (as a measure of coherence). Our results demonstrate that entanglement is highly sensitive to the scattering geometry and disappears near right-angle scattering, while quantum coherence remains finite even in regimes where entanglement vanishes completely.
These findings provide a unified description of polarization-dependent decoherence in annihilation photon pairs and clarify the distinct roles of entanglement and coherence in realistic two-photon interactions.
These results are relevant for quantum-enhanced positron emission tomography and highlight the persistence of quantum resources in scattering-dominated media.


\end{abstract}


\section{Introduction}

Electron-positron annihilation is a fundamental process in quantum electrodynamics (QED), producing entangled pairs of 511 keV photons. 
This process can occur either directly through free positron annihilation or indirectly via positronium (Ps) formation at rest. When a positron slows down in matter, it may form a short-lived bound state with an electron, known as positronium, before the annihilation event \cite{Bass}. Positronium exists in two spin configurations: para-positronium (p-Ps), a singlet state with antiparallel spins (S=0), and ortho-positronium (o-Ps), a triplet state with parallel spins (S=1) \cite{Bass, Hiesmayr, Moskal2019}. The p-Ps predominantly decays into an entangled photon pair, whereas o-Ps decays into three photons \cite{Bass, Hiesmayr, Moskal2019, Acin, Nowakowski}.
Experimental observations confirm that positronium formation and photon correlations depend strongly on the properties of the matter, as demonstrated in recent positron emission tomography (PET) studies using plastic J-PET scanners \cite{Moskal2025}.

One of the most remarkable features of quantum mechanics is the existence of nonlocal correlations between entangled particles that cannot be explained within the framework of classical physics. This phenomenon was first introduced in the well-known Einstein-Podolsky-Rosen (EPR) paradox, which challenged the completeness of quantum mechanics \cite{EPR}. Later, Bohm and Aharonov reformulated this concept and emphasized its physical significance \cite{Bohm}. Nonlocal correlations, together with the violation of Bell's inequalities, have demonstrated that nature exhibits fundamentally nonclassical behavior \cite{Bell,Clauser,Aspect}.
Quantum correlations between photons can persist over large propagation distances and times if decoherence due to the surrounding medium is minimized \cite{berera1,berera2}. For annihilation photons, such correlations arise directly from conservation laws and offer a natural framework for studying photon-matter interactions. 

Beyond their theoretical importance, annihilation photons are emerging as promising tools for quantum imaging and quantum-enhanced PET \cite{Zaidi2007,Mcnamara,Moskal2018,Toghyani,Caradonna2019,Watts}.
Their strong correlations enable applications such as timing calibration, detector characterization, and coincidence-based gamma-ray spectroscopy \cite{Moskal2018}. Preserving coherence and entanglement between their polarization states allows deeper insight into the quantum nature of photon-matter interactions.

Both entanglement and quantum coherence are sensitive to how much classical information about the system becomes accessible. For example, when a photon undergoes Compton scattering, interaction with an external electron leads to a gradual, scattering-angle-dependent reduction of entanglement \cite{caradonna3}. Similarly, in quantum double-double-slit experiments with momentum-entangled photons, revealing which-slit path information for either photon collapses entanglement and suppresses single-photon interference, while the two-photon interference pattern persists only when the path information remains hidden \cite{kaur}. These phenomena can be described by medium-induced disentanglement \cite{zurek}, in which the transfer of quantum information into accessible classical channels leads to the selection of pointer states and the suppression of quantum superpositions. Consequently, when a classical property of one particle in an entangled pair (such as path, momentum, spin, or polarization) becomes fully distinguishable, the effective entanglement of the system vanishes.

Despite extensive research on photon entanglement and its applications in PET imaging, the quantitative behavior of entanglement and coherence under multiple Compton scattering processes remains insufficiently explored. In realistic detection media, such as biological tissues or detector materials, annihilation photons may undergo one or more Compton scattering events before detection. These interactions can alter photon polarizations and correlations, affecting image reconstruction, the signal-to-noise ratio, and the reliability of coincidence-based measurements in PET techniques. Therefore, a detailed theoretical analysis of how Compton scattering in a medium influences quantum entanglement and coherence is essential. In particular, the relative behavior of entanglement and quantum coherence under consecutive Compton scattering events has not been quantitatively investigated.

Previous studies, particularly those by Caradonna \textit{et al.} \cite{caradonna3,caradonna2}, have shown that Compton scattering does not necessarily destroy quantum entanglement; under certain conditions, polarization correlations can survive even after one photon scatters. However, a comprehensive quantitative comparison of entanglement and coherence under single, double, and triple scattering processes has not been reported. The present study aims to fill this gap by providing a framework for analyzing the evolution of quantum entanglement and coherence in the annihilation photon pair under consecutive Compton scattering events. Using Stokes matrix formalism \cite{caradonna1,caradonna2025} and density matrix techniques, we model the complete polarization dynamics of the two-photon system and compute two quantitative measures: concurrence as a measure of entanglement and the $l_1$-norm of coherence as an indicator of quantum superposition.

Our results reveal that both concurrence and the $l_1$-norm strongly depend on the scattering geometry, often exhibiting opposite behaviors: quantum coherence can persist even when entanglement is substantially reduced. 
Recent studies in quantum resource theory highlight a quantitative link between entanglement and coherence: local coherence can be converted into bipartite entanglement via incoherent operations, and conversely, entanglement can generate coherence in subsystems \cite{Streltsov2015,Chitambar2016,Zhu2017}.
These findings underscore the importance of analyzing both resources simultaneously, particularly in realistic scattering scenarios, since coherence often proves to be a more robust and long-lived quantum resource than entanglement under non-unitary processes.
This distinction provides new insight into the resilience of quantum properties in practical settings. A similar behavior of quantum coherence and entanglement has also been reported in neutrino oscillation systems \cite{ettefaghi1,ettefaghi2}.

Unlike previous studies that focused primarily on entanglement under single Compton scattering, the present work provides a systematic and comparative analysis of both entanglement and quantum coherence under consecutive scattering events. This approach identifies scattering geometries in which nonlocal correlations are suppressed while local polarization coherence remains significant, a feature that has not previously been explored quantitatively. Such insights are particularly relevant for applications in quantum-enhanced PET imaging and for understanding the persistence of quantum resources in scattering-dominated media. 

In the following sections, we first formulate the quantum state of entangled photon pairs and derive the final density matrix after one, two, and three Compton scattering events using the Stokes matrix formalism. Sections \ref{three} and \ref{four} analyze the variation of entanglement and quantum coherence under one, two, and three Compton scattering processes. Finally, in the Discussion section, we analyze the implications of our findings, and in the conclusion, we summarize the main results and outline their potential applications.

\section{ Density Matrix of the Entangled Photon Pair State}\label{section2}

In quantum physics and optics, several theoretical frameworks are used to describe the behavior of photons, each suited to a particular aspect of their dynamics. Wave packets, for example, are widely used in studies involving spatial or temporal coherence and interferometric propagation \cite{Mandel1995, Bialynicki1996, Smith2007}. In this work, we focus on polarization correlations, for which the Stokes-Mueller formalism provides a natural and powerful description. This framework is especially suited for modeling photon-environment interactions, depolarization, and other nonunitary processes such as Compton scattering, which play a central role in positron annihilation imaging techniques \cite{ Goldberg2021}.

We begin by considering the two-photon state produced in p-Ps annihilation at rest. In the center-of-mass frame, the electron and positron annihilate into back-to-back photons with equal energies. Under ideal conditions, the polarization state of the two-photon is described by the maximally entangled Bell singlet state \footnote{The four Bell states form a complete basis of maximally entangled two-qubit states:
$|\Psi^{\pm}\rangle = \frac{1}{\sqrt{2}} \left( |H_1 V_2\rangle \pm |V_1 H_2\rangle \right)$ and $|\Phi^{\pm}\rangle = \frac{1}{\sqrt{2}} \left( |H_1 H_2\rangle\pm |V_1 V_2\rangle \right)$ }
\begin{equation}
|\Psi^{-}\rangle = \frac{1}{\sqrt{2}} \left( |H_1 V_2\rangle - |V_1 H_2\rangle \right),\label{state}
\end{equation}
where $|H\rangle$ and $|V\rangle$ denote horizontal and vertical polarizations, and the subscripts 1 and 2 refer to the first and second photons, respectively. Due to Bose symmetry and parity conservation in the decay of p-Ps, the two-photon system therefore forms a spin-singlet state and is maximally entangled in polarization \cite{Hiesmayr2019}. 

The corresponding density matrix is
 \begin{equation}
\rho=|\Psi^-\rangle\langle\Psi^-|,
 \end{equation}
 which represents a pure quantum state satisfying $\mathrm{Tr}(\rho^2)=1$. In the ordered basis $|HH\rangle, |HV\rangle, |VH\rangle$, and $|VV\rangle$, this matrix takes the explicit form
 \begin{equation}
 	\rho=\frac{1}{2}
 	\begin{pmatrix}
 		0 & 0 & 0 & 0 \\
 		0 & 1 & -1 & 0 \\
 		0 & -1 & 1 & 0 \\
 		0 & 0 & 0 & 0
 	\end{pmatrix}.
 \end{equation}
 
To describe how Compton scattering modifies this two-photon polarization state, we use the generalized Stokes formalism for bipartite systems \cite{caradonna2,caradonna1,caradonna2025}. 
The two-photon Stokes matrix is defined as
\begin{equation}
	S_{ij} = \operatorname{Tr}\left[ \rho \, (\sigma_i \otimes \sigma_j) \right],
\end{equation}
where $\sigma_0 = \mathbb{I}$ and $\sigma_{1,2,3}$ are the Pauli matrices. Substituting the initial state yields the diagonal matrix
\begin{equation}
	S =
	\begin{pmatrix}
		1 & 0 & 0 & 0 \\
		0 & -1 & 0 & 0 \\
		0 & 0 & -1 & 0 \\
		0 & 0 & 0 & -1
	\end{pmatrix},
\end{equation}
which reflects the perfect anti-correlation of all three polarization components in the singlet state.

When one or both photons undergo Compton scattering, their Stokes parameters evolve according to the Mueller matrices $T_{pk}$, describing the scattering process, and the rotation matrices $M_{pk}$ that account for the change of polarization basis due to the scattering geometry (see Appendix \ref{AppendixA}).

The polarization-dependent Compton scattering formalism employed here is derived from the Klein-Nishina cross section \cite{klein1929}, which allows the derivation of the Stokes matrix \cite{Fano,Mcmaster}. This $4\times4$ matrix provides a complete QED-consistent description of the polarization transformation during a single Compton scattering event.
In this work, we employ this matrix to construct the density matrix of two entangled annihilation photons undergoing successive scatterings, enabling analytical evaluation of concurrence and the $l_1$-norm.

 Following the approach of \cite{caradonna2025}, the polarization basis is first rotated into the local scattering plane, after which the Compton Mueller matrix is applied. For multiple scattering events, these transformations are applied sequentially in the chronological order of interactions.

The evolved two-photon Stokes matrix after $m$ scatterings on photon 1 and $n$ scatterings on photon 2 is given by
 \begin{equation}
S'=\Big(\prod_{k=m}^1T_{1k} M_{1k}\Big) S \Big(\prod_{k=n}^1T_{2k}M_{2k} \Big)^\top.
 \end{equation}

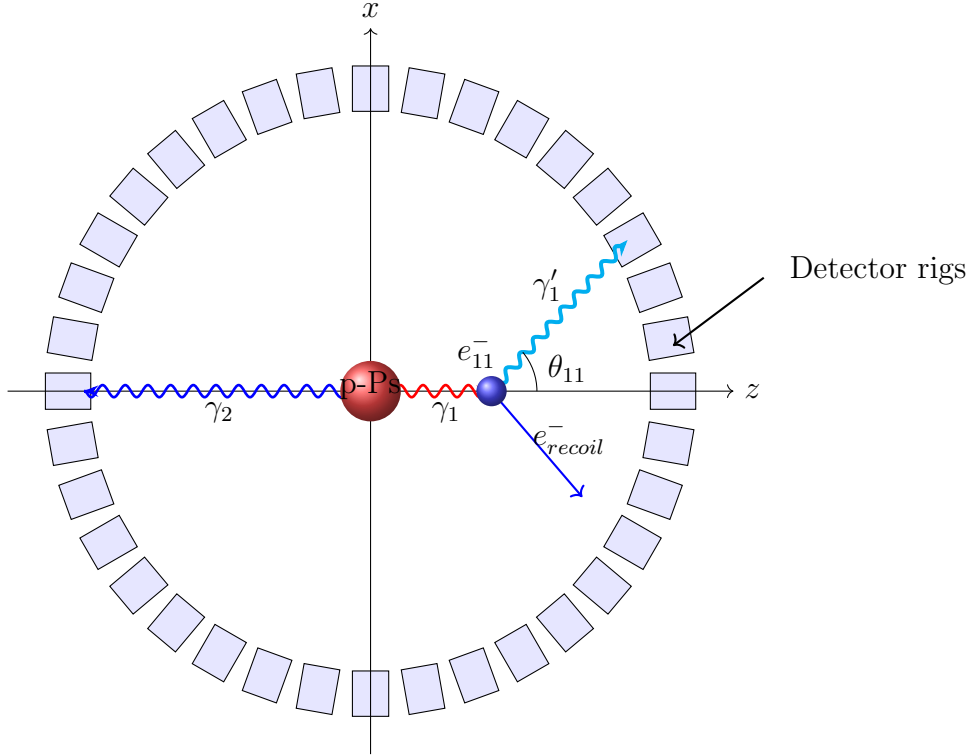
\begin{figure}[ht]
	\centering
	\begin{tikzpicture}[scale=2]
	
	\def\R{2}
	
	\foreach \a in {0,10,...,355}{
		\begin{scope}[shift={(\a:\R)}, rotate=\a]
		\draw[fill=blue!10] (-0.15,0.12) rectangle (0.15,-0.12);
		\end{scope}	}
	
	\draw[->] (-2.4,0)--(2.4,0) node[right]{$z$};
	\draw[->] (0,-2.4)--(0,2.4) node[above]{$x$};
	
	\draw[-{Stealth[length=2mm]},thick,red,line width=1pt,
	decorate,decoration={snake,amplitude=0.8mm,segment length=3mm}]
	(0,0,0) -- (0.8,0,0);
	\node[below] at (0.5,0,0) {$\gamma_1$};
	
	\draw[-{Stealth[length=2mm]},thick,blue,line width=1pt,
	decorate,decoration={snake,amplitude=0.8mm,segment length=3mm}]
	(0,0,0) -- (-1.9,0,0);
	\node[below] at (-1,0,0) {$\gamma_2$};
	
	\shade[ball color=red!70, tdplot_screen_coords] (0,0,0) circle (0.2);
	\node[above] at (-0.08,-0.2,-0.2) {p-Ps};
	
	\draw[-{Stealth[length=2.2mm]},very thick,cyan,line width=1.5pt,
	decorate,decoration={snake,amplitude=0.6mm,segment length=2.7mm}]
	(0.8,0)--(1.7,1);
	
	\node[right] at (1,0.7) {$\gamma'_1$};
	
	\draw[blue,->,thick] (0.8,0)--(1.4,-0.7);
	\node[right] at (1,-0.3) {$e^{-}_{recoil}$};
	
	\shade[ball color=blue!70, tdplot_screen_coords] (0.8,0) circle (0.1);
	\node[above] at (0.7,0.08) {$e^-_{11}$};
	
	\draw (1.1,0) arc (0:40:0.4);
	\node at (1.3,0.15) {$\theta_{11}$};

	\node[right] at (2.7,0.8) {Detector rigs};
	\draw[->,thick] (2.6,0.75) -- (2.0,0.3);

	\end{tikzpicture}
	\caption{\footnotesize{Schematic of the single Compton scattering geometry for one of the annihilation photons. The back-to-back photons $\gamma_1$ and $\gamma_2$ originate from para-positronium (p-Ps) decay at the center of a PET-like detector ring. Photon $\gamma_1$ scatters off an electron $e^-_{11}$ at polar angle $\theta_{11}$, producing a scattered photon $\gamma'_1$ and a recoil electron. The azimuthal angle $\phi_{11}$, which defines the rotation of the scattering plane around the incident photon, is omitted for clarity. Polarization correlation measures (concurrence and $l_1$-norm) are evaluated in subsequent figures as functions of $\theta$ and $\phi$; here, only the geometry of a single scattering event is shown.}}
	\label{schematic}
\end{figure}

For clarity, we consider the three scenarios used throughout this work:
 \begin{enumerate}
	\item{Single Compton scattering:} one scattering on photon 2 
	\begin{equation}
		S'= S \,M_{21}^\top T_{21}^\top,\label{single}
	\end{equation}
	\item{Double Compton scattering:} one scattering on each photon
	\begin{equation}
		S'= T_{11}M_{11} \,S\,M_{21}^\top T_{21}^\top,\label{double}
	\end{equation}
	\item{Triple Compton scattering:} one scattering on photon 1 and two consecutive scatterings on photon 2
	\begin{equation}
		S'= T_{11}M_{11} \,S\,M_{21}^\top T_{21}^\top M_{22}^\top T_{22}^\top.\label{triple}
	\end{equation}
\end{enumerate}
A schematic representation of the single Compton scattering scenario is shown in Fig. \ref{schematic}, illustrating the geometry of photon emission, scattering, and detection. The extension to multiple scattering events follows analogous principles.

Once the final Stokes matrix $S'$ is obtained, the associated density matrix is reconstructed as \cite{caradonna3,abouraddy}
 \begin{equation}\label{density matrix}
 	\rho'=\frac{\sum_{i,j=0}^{3}S'_{ij}\sigma_i\otimes \sigma_j}{\mathrm{Tr}[\sum_{i,j=0}^{3}S'_{ij}\sigma_i\otimes \sigma_j]},
 \end{equation}
which guarantees a Hermitian and normalized density operator. In Appendix \ref{AppendixB}, we explicitly computed $\rho'$ for the single scattering case (Eq. \ref{dm}). The expressions for the double and triple scattering scenarios are omitted for brevity due to their length. 

The reconstructed density matrix serves as the basis for quantifying how successive Compton interactions modify the entanglement and coherence of the annihilation photon pair. In the following sections, we analyze these quantum resources under single, double, and triple scattering scenarios.
 
 All numerical evaluations of the two-photon density matrices, as well as the computation of concurrence and the $l_1$-norm, were performed in Python (version 3.11) using NumPy and SciPy for linear-algebra operations and Matplotlib for plotting.

The initial photons originate from p-Ps annihilation at rest with energy $E_{0} = 511 \text{keV}$. In the MeV regime, the scattered photon energy is determined solely by the scattering geometry through the standard Compton formula in Eq. \ref{16e}, which relates the scattered energy to the scattering angle $\theta$.
	
The scattering geometries employed in the figures correspond to configurations readily achievable in laboratory and PET-based measurements. The specific scattering and azimuthal angles used for each simulation are listed explicitly in the figure captions.

 \section{Quantification of Entanglement Using Concurrence Measure}\label{three}
 Quantum entanglement in the annihilation photon pair arises from strong correlations between their polarization degrees of freedom. In p-Ps annihilation at rest, the two emitted photons form a spin-singlet Bell state, such that a polarization measurement on one photon immediately fixes the polarization outcome of the other, regardless of their spatial separation. These nonclassical correlations are naturally characterized using entanglement measures such as concurrence \cite{Streltsov2015,Chitambar2016,Zhu2017}, which provide a direct framework for quantifying how environmental interactions (such as Compton scattering) progressively degrade quantum correlations.
 
In this work, we quantify the entanglement of the annihilation photon pair using the concurrence measure introduced by Wootters for two-qubit systems \cite{hillwootters1997,wootters1998}. 
Concurrence is computed directly from the two-qubit density matrix and is therefore well suited for analyzing situations in which an initially pure Bell state evolves into a mixed state through coupling with an environment \cite{vedral1997,horodecki2009}. 
For a general two-qubit density matrix $\rho$, the concurrence is defined as
\begin{equation}
	C(\rho)= \max\left(0,\sqrt{\lambda_1} - \sqrt{\lambda_2} - \sqrt{\lambda_3} - \sqrt{\lambda_4}\right),
\end{equation}
where $\lambda_i$ ($i=1,2,3,4$) are the eigenvalues, in decreasing order, of the non-Hermitian matrix $\rho\,\tilde{\rho}$ with 
\begin{equation}
\tilde{\rho}= (\sigma_y \otimes \sigma_y)\,\rho^*\,(\sigma_y \otimes \sigma_y),
\end{equation}
and $\rho^*$ denotes complex conjugation of $\rho$ in the chosen computational basis.  
Concurrence ranges from $C(\rho)=0$ for separable states (non-entangled) to $C(\rho)=1$ for maximally entangled states; in particular, the initial Bell state $|\Psi^{-}\rangle$ in Eq.\eqref{state} has concurrence equal to one.

The effect of Compton scattering on the two-photon state is incorporated using the generalized Stokes-Mueller formalism described in Sec.\ref{section2}. After each scattering event, the Stokes matrix of the system is transformed, and the final density matrix $\rho'$ is reconstructed via Eq. \ref{density matrix}. Throughout this work, the scattering electrons are assumed to be initially at rest; more general scenarios, involving moving electrons, can be modeled through Lorentz transformations of the scattering geometry. Evaluating the concurrence of $\rho'$ quantifies the surviving entanglement after a given sequence of scattering events. Since Compton scattering is a non-unitary process, part of the polarization information becomes encoded in inaccessible environmental degrees of freedom, leading to a reduction of concurrence.

Our numerical results indicate that the number of scattering events and the polar scattering angles play the dominant roles in entanglement degradation, while azimuthal angles introduce only secondary modulations.
Under strong depolarization (such as multiple successive scatterings or large scattering angles in a dense medium), the two-photon state approaches a maximally mixed state, and the concurrence tends to zero \cite{vedral1997, YuEberly}.
To quantify these effects in detail, we compute concurrence for single, double, and triple Compton scattering events (see the subsections below). These calculations allow us to track the rate at which entanglement is lost as the environment progressively acquires polarization information.

 \subsection{Single Compton Scattering}

To evaluate the effect of a single Compton scattering event on polarization entanglement, we consider the scenario in which only photon 2 undergoes one scattering event with an electron initially at rest, while the partner photon propagates freely. The two-photon Stokes matrix evolves according to Eq. \ref{single},
and the corresponding final density matrix $\rho'$ is constructed via Eq.\ref{density matrix}. A closed-form expression for the concurrence in this configuration is provided in Appendix \ref{AppendixB} (Eq. \ref{con}).

Figure \ref{concurrence13d} shows the concurrence as a function of the scattering angles $\theta_{21}$ and $\phi_{21}$. The numerical results reveal a monotonic decrease in concurrence as $\theta_{21}$ approaches $90^\circ$, reaching a near zero-value within the narrow interval $85^\circ\lesssim \theta_{21}\lesssim96.8^\circ$, in agreement with previous findings \cite{caradonna3}. This loss of entanglement near right-angle scattering reflects the strong coupling between the photon polarization and the scattering electron in this angular range; the polarization effectively becomes which-polarization information in the environment, and is therefore lost from the two-photon subsystem.
For larger scattering angles ($\theta_{21}>97^\circ$), the concurrence increases again, indicating that some polarization correlations survive in specific geometries.

\begin{figure}[h!]
	\centering
	\begin{subfigure}[b]{0.49\textwidth}
		\centering
		\includegraphics[width=\textwidth]{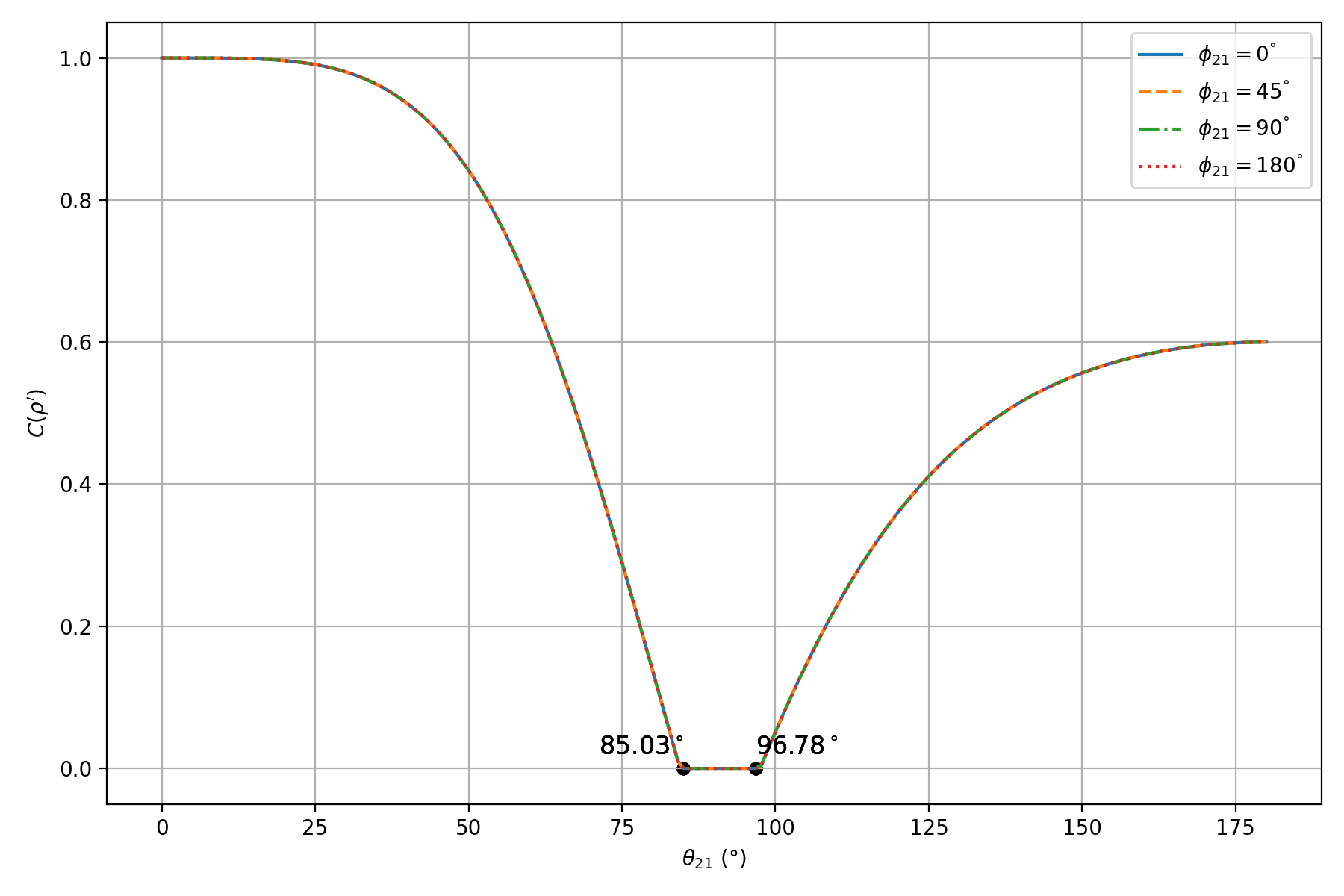}
	\end{subfigure}
	\hfill
	\begin{subfigure}[b]{0.49\textwidth}
		\centering
		\includegraphics[width=\textwidth]{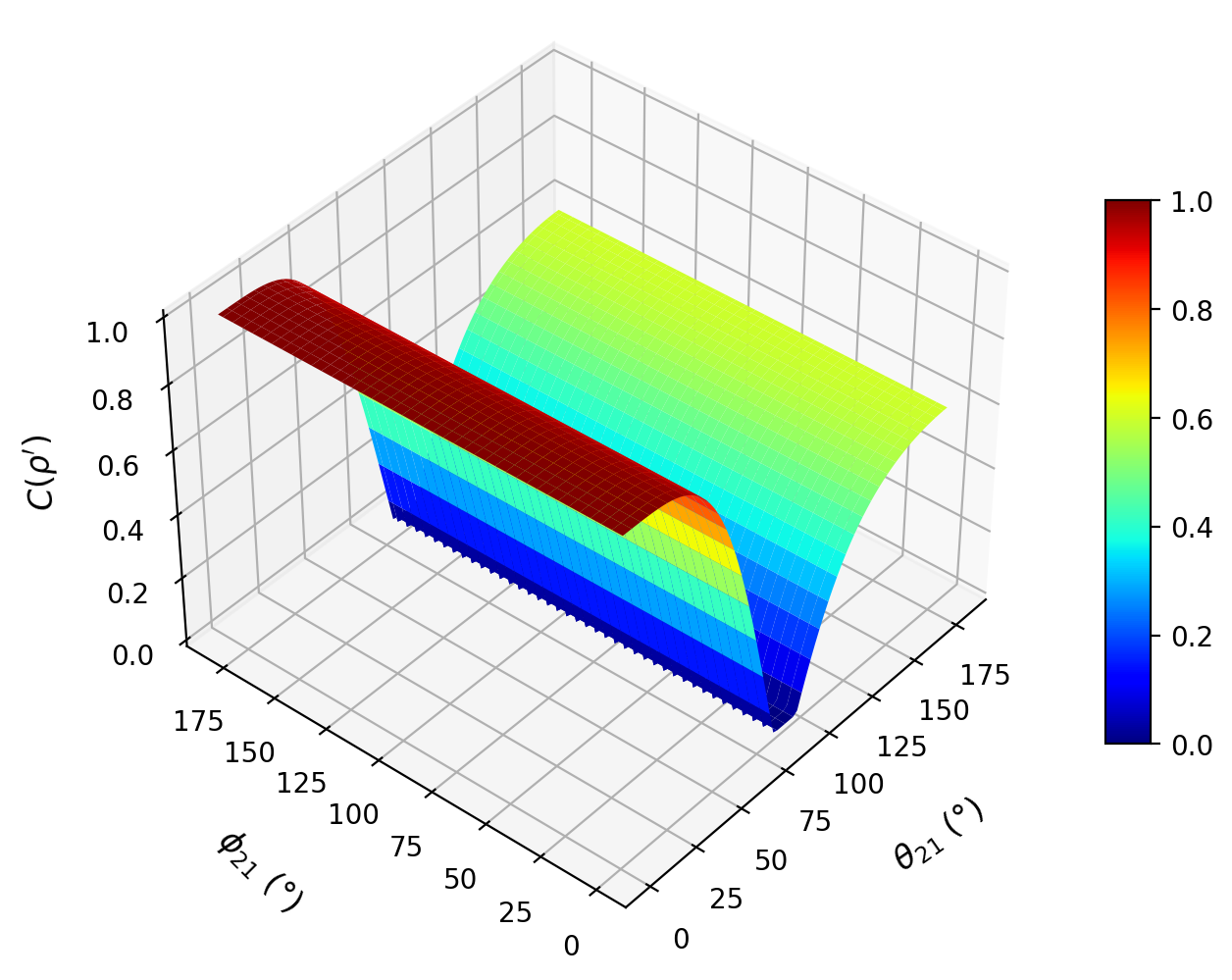}
	\end{subfigure}
	\caption{\footnotesize Concurrence $C(\rho')$ for single Compton scattering. (Left) 2D dependence on polar angle $\theta_{21}$, and (Right) 3D plot as a function of $\theta_{21}$ and $\phi_{21}$. Concurrence depends only on $\theta_{21}$ and exhibits a sharp minimum near right-angle scattering.}
	\label{concurrence13d}
\end{figure}

Importantly, the concurrence in the single scattering case is independent of the azimuthal angle $\phi_{21}$ (Fig. \ref{concurrence13d}). Variations in $\phi_{21}$ correspond only to local unitary rotations of the polarization basis around the scattering axis and therefore do not affect entanglement.
 
  \begin{figure}[h!]
 	\centering
 	\centering
 	\includegraphics[width=0.6\textwidth]{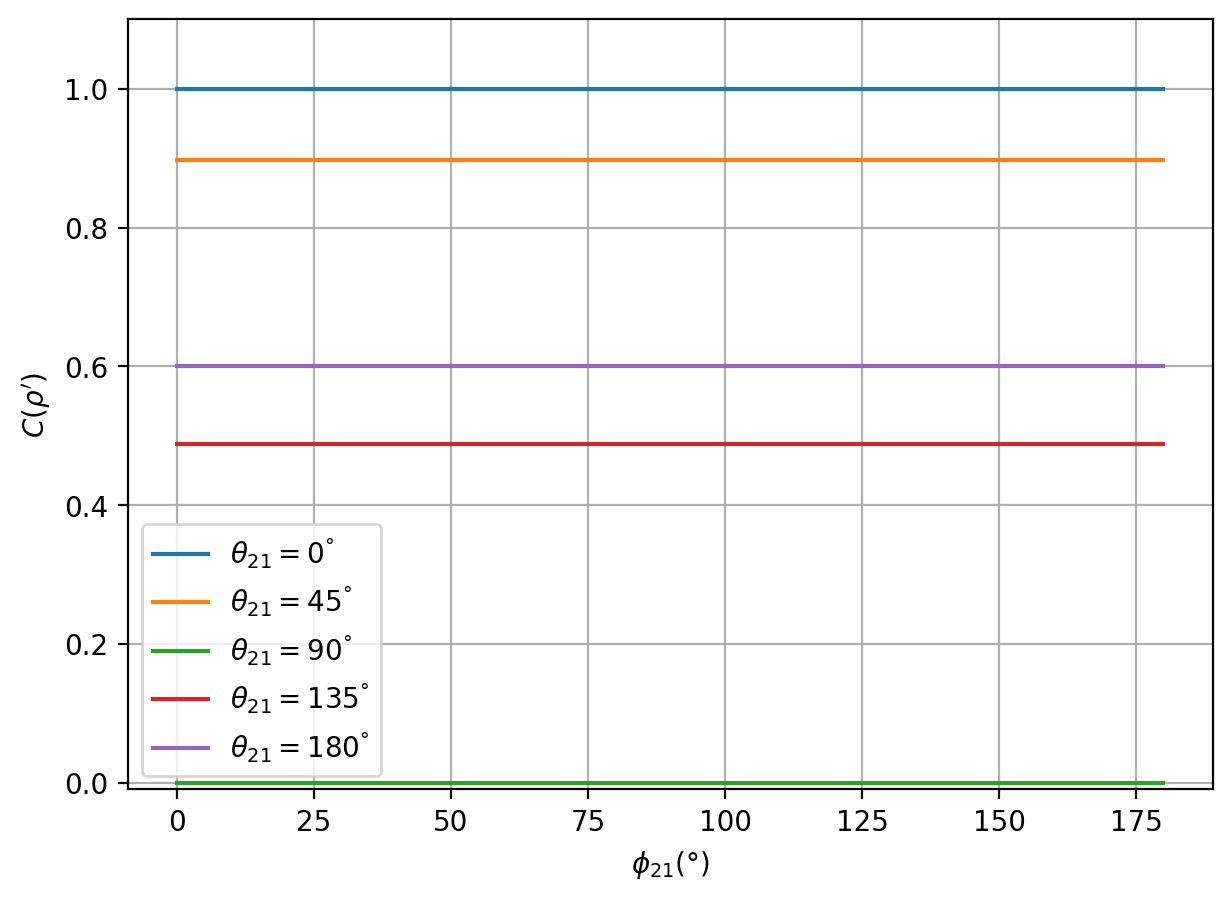}
 	\caption{\footnotesize Concurrence $C(\rho')$ for single Compton scattering as a function of azimuthal angle $\phi_{21}$ for several fixed polar angles. All curves appear as horizontal lines, confirming that $C(\rho')$ remains invariant under changes in $\phi_{21}$.}
 	\label{concurrence1phi}
 \end{figure}

 \subsection{Double Compton Scattering} 
 
 In the double scattering configuration, each photon undergoes one Compton scattering. The corresponding Stokes matrix is obtained using Eq. \ref{double}. 
The final density matrix depends on two polar angles ($\theta_{11}$ and $\theta_{21}$) and on the relative azimuthal angle difference ($\Delta\phi=\phi_{21}-\phi_{11}$), which together determine the full scattering geometry.

\begin{figure}[h!]
	\centering
	\begin{subfigure}{0.49\textwidth}
		\centering
		\includegraphics[width=\textwidth]{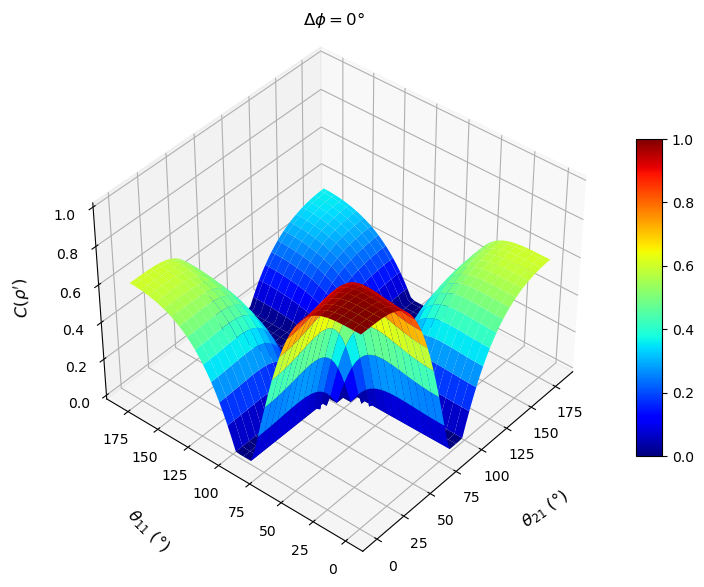}
	\end{subfigure}
	\hfill
	\begin{subfigure}{0.49\textwidth}
		\centering
		\includegraphics[width=\textwidth]{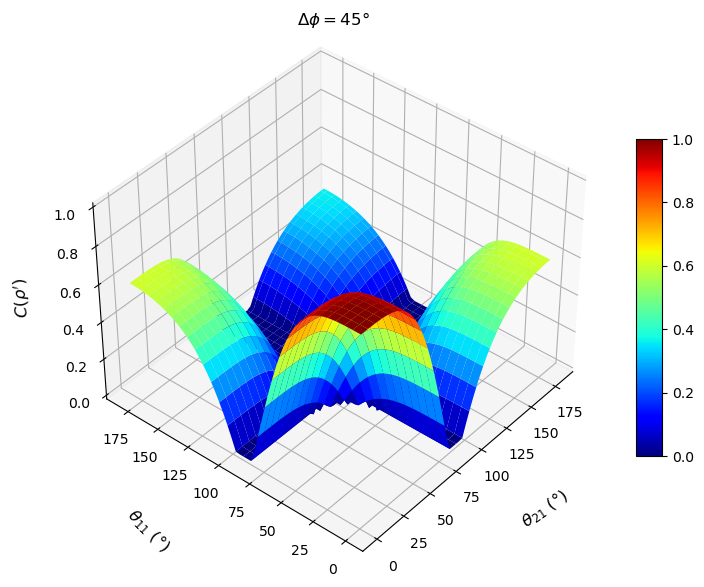}
	\end{subfigure}
	\hfill
	\begin{subfigure}{0.49\textwidth}
		\centering
		\includegraphics[width=\textwidth]{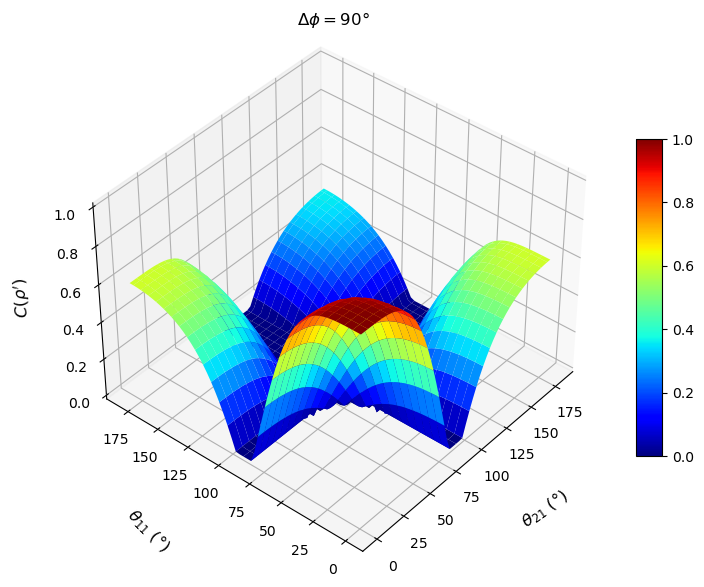}
	\end{subfigure}
	\hfill
	\begin{subfigure}{0.49\textwidth}
		\centering
		\includegraphics[width=\textwidth]{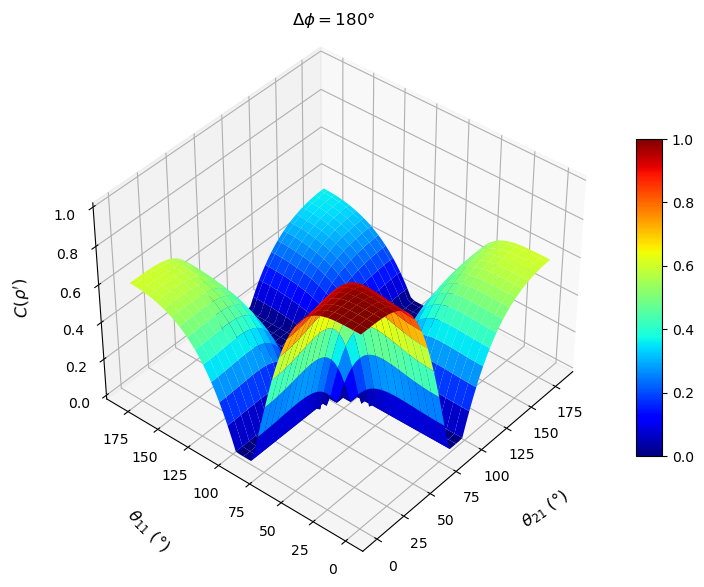}
	\end{subfigure}
	\caption{\footnotesize Concurrence $C(\rho')$ for double Compton scattering as a function of the two polar angles $\theta_{11}$ and $\theta_{21}$ for several azimuthal differences $\Delta\phi$. A broad region of complete disentanglement emerges near right-angle scattering.}
	\label{concurrence3d}
\end{figure}

Figure \ref{concurrence3d} shows the concurrence surface across the $(\theta_{11},\theta_{21})$ plane for several values of $\Delta\phi$. The results reveal a symmetry under the exchange of the two polar angles and exhibit a pronounced valley of zero-concurrence when either or both of the polar angles lie near $90^\circ$. This reflects the fact that each scattering event independently transfers polarization information to the environment; when either or both photons scatter in polarization-sensitive directions, the entanglement is effectively destroyed.

\begin{figure}[h!]
 	\centering
 	\includegraphics[width=0.6\textwidth]{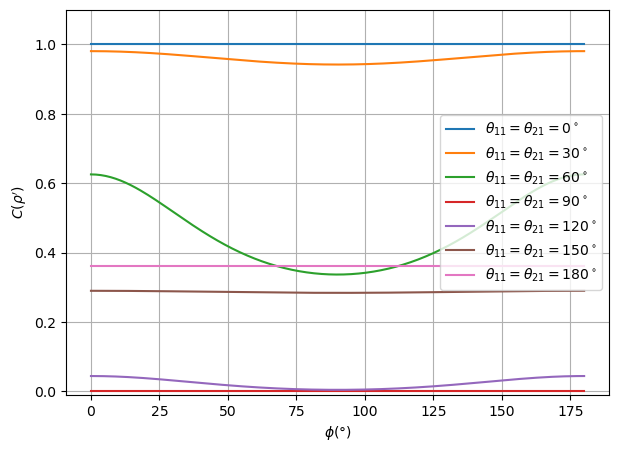}
 	\caption{\footnotesize Concurrence $C(\rho')$ for double Compton scattering as a function of the azimuthal angle $\Delta\phi$ for several fixed polar angles. The dependence on $\Delta\phi$ becomes noticeable only for $\theta\gtrsim30^\circ$, while for $\theta\approx90^\circ$ the concurrence vanishes for all azimuthal configurations.}
 	\label{concurrence2phi}
 \end{figure}

In contrast to the single scattering case, the concurrence here exhibits a nontrivial dependence on the azimuthal angle $\Delta\phi$, especially for intermediate polar angles (Fig. \ref{concurrence2phi}). For small polar angles ($\theta\lesssim30^\circ$, assuming $\theta=\theta_{11}=\theta_{21}$), the dependence on $\Delta\phi$ is negligible. As $\theta$ increases, the sensitivity to $\Delta\phi$ becomes more pronounced, but near $\theta=90^\circ$ concurrence vanishes for all azimuthal configurations. For backscattering angles beyond the zero-entanglement region, concurrence may reappear but remains typically small and approximately independent of $\Delta\phi$.

 \subsection{Triple Compton Scattering}
 The triple scattering case consists of one scattering on photon 1 and two consecutive scatterings on photon 2. The Stokes matrix obtained using Eq. \ref{triple},
and the reconstructed density matrix depends on three polar angles ($\theta_{11}, \theta_{21}, \theta_{22}$) and the corresponding azimuthal angles ($\Delta\phi, \phi_{22}$).
 
\begin{figure}[h!]
	\centering
	\begin{subfigure}{0.49\textwidth}
		\centering
		\includegraphics[width=\textwidth]{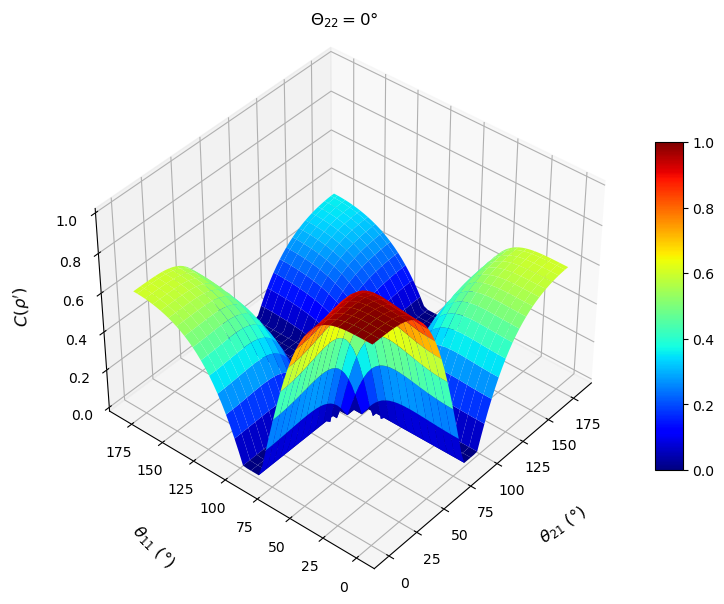}
	\end{subfigure}
	\hfill
	\begin{subfigure}{0.49\textwidth}
		\centering
		\includegraphics[width=\textwidth]{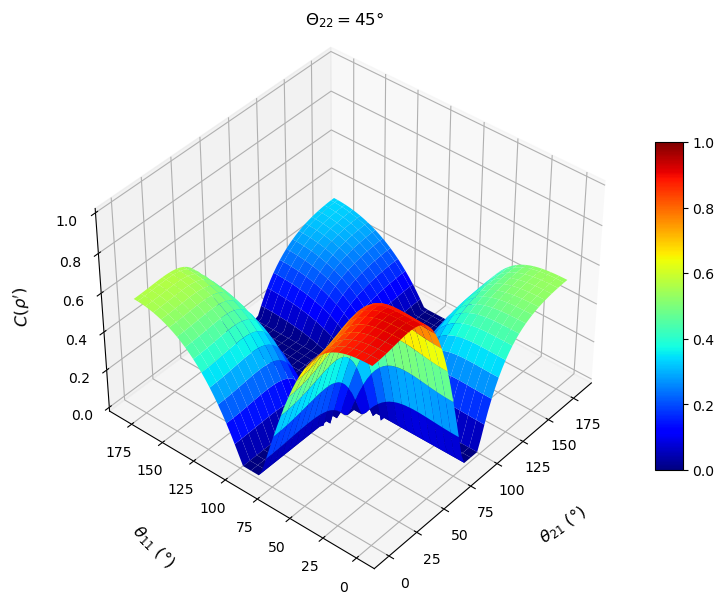}
	\end{subfigure}
	\hfill
	\begin{subfigure}{0.49\textwidth}
		\centering
		\includegraphics[width=\textwidth]{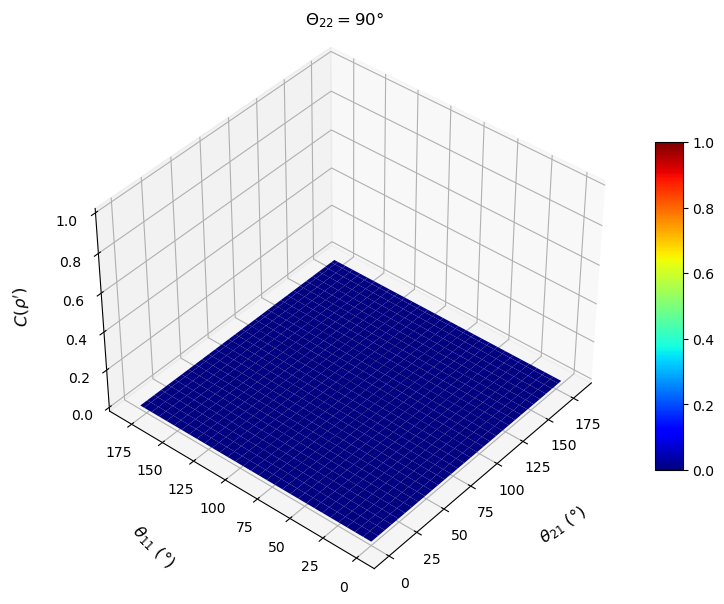}
	\end{subfigure}
	\hfill
	\begin{subfigure}{0.49\textwidth}
		\centering
		\includegraphics[width=\textwidth]{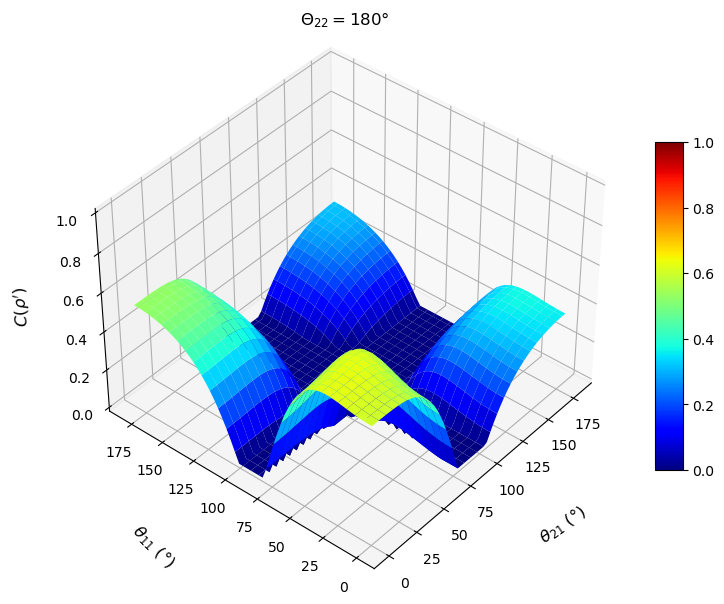}
	\end{subfigure}
	\caption{\footnotesize Concurrence $C(\rho')$ for triple Compton scattering as a function of $\theta_{11}$ and $\theta_{21}$ for several values of $\theta_{22}$. Larger values of $\theta_{22}$ suppress entanglement and broaden the zero-concurrence region.}
	\label{concurrence33d}
\end{figure}

The concurrence surfaces (Fig. \ref{concurrence33d}) show that the second scattering on photon 2 substantially enhances entanglement degradation. As $\theta_{22}$ increases, the maximum achievable concurrence decreases, and the zero-entanglement region expands significantly.
When either or both of the polar angles approach $90^\circ$, concurrence collapses; the effect is most pronounced when $\theta_{22}\approx90^\circ$, where the disentanglement valley extends over a wide area in the $\theta_{11}$ and $\theta_{21}$ plane. 
This cumulative degradation is consistent with the interpretation of successive non-unitary interactions progressively leaking polarization information to the environment.

\begin{figure}[h!]
	\centering
	\includegraphics[width=0.6\textwidth]{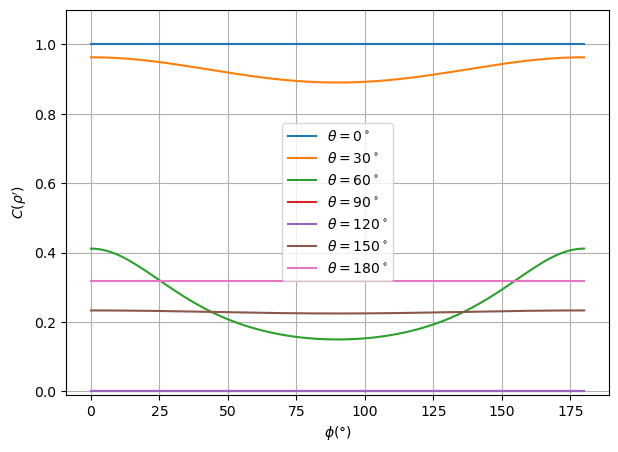}
	\caption{\footnotesize Concurrence $C(\rho')$ for triple Compton scattering as a function of $\Delta\phi$. Azimuthal variations affect the concurrence away from the zero-entanglement region, but the location of the zero-concurrence interval is determined by the polar angles.}
	\label{concurrence3phi}
\end{figure}

Azimuthal variations modulate the concurrence only away from the disentanglement regions and do not shift the central location of the zero-concurrence region, which remains governed by the polar scattering angles (Fig. \ref{concurrence3phi}).

 \section{Quantification of Quantum Coherence Using the $l_1$-norm Measure}\label{four}
Quantum coherence quantifies the ability of a quantum system to exist in a superposition of basis states. For annihilation photons, this appears as a coherent superposition of linear polarization states. Unlike entanglement, which characterizes intrinsically nonlocal correlations between spatially separated photons, quantum coherence can exist both locally within each photon and globally across the composite two-photon system. These two resources are deeply connected: global coherence is a necessary prerequisite for bipartite entanglement, and coherence can be converted into entanglement under incoherent operations assisted by classical communication \cite{Streltsov2015,Chitambar2016}. Consequently, when a physical process suppresses nonlocal correlations and destroys entanglement, residual local coherence may still persist within the photon subsystems \cite{Zhu2017}.
From the perspective of quantum resource theory, coherence constitutes a fundamental quantum resource underlying many quantum technologies, including quantum computation, communication, and metrology \cite{baumgratz2014,streltsov2017}.  

The $l_1$-norm of the density matrix provides a simple, physically transparent, and experimentally accessible measure of polarization coherence \cite{Streltsov2015,Zhu2017,Singh2015}.
This measure, introduced by Baumgratz \textit{et al.} \cite{baumgratz2014} for a density matrix $\rho'$, is defined as
\begin{equation}
	C_{l_1}(\rho) = \sum_{i\neq j} |\rho_{ij}|,
\end{equation}
where the sum runs over all off-diagonal elements in the chosen basis.
For a $d$-dimensional system, the $l_1$-norm is bounded above by $d-1$ (achieved by the maximally coherent pure state), so for a two-qubit system, the maximum possible value is 3.
A value of zero indicates a fully incoherent state (diagonal density matrix), while larger values quantify stronger polarization superposition. This measure is particularly suitable for both theoretical and experimental studies due to its straightforward computation and direct physical interpretation \cite{streltsov2017, Singh2015}.

In p-Ps annihilation, the two-photon begin in a maximally coherent Bell state in the linear polarization basis. This coherence directly reflects the magnitude of off-diagonal elements in the density matrix, which encode the well-defined superposition of polarization states. Compton scattering introduces decoherence by coupling photon polarization to environmental electrons, reducing the magnitude of the off-diagonal terms in the density matrix. Using the generalized Stokes-Mueller formalism (Sec. 2), we reconstruct the evolved density matrix $\rho'$ after each scattering event and compute the corresponding coherence.

To examine how coherence decays under realistic interaction two-photon, we evaluate the $l_1$-norm for three scenarios: single, double, and triple scattering events. This also allows direct comparison with entanglement (Sec. 3), revealing their different sensitivities to scattering geometry and the non-unitary character of Compton interactions.

Beyond foundational significance, the study of $l_1$-norm coherence in annihilation photons is also of practical importance for positron emission tomography (PET) imaging. The partial preservation of polarization coherence after scattering can influence coincidence detection, timing resolution, and the contribution of scattered photons to image noise. Understanding how coherence degrades through multiple scattering events reveals valuable guidance for optimizing detector materials and geometries in advanced and quantum-enhanced PET imaging \cite{Caradonna2019,baumgratz2014,streltsov2017}.

Finally, the behavior of coherence provides a rigorous and experimentally accessible framework for quantifying scattering-induced decoherence in the annihilation photon pair. Evaluating $C_{l_1}(\rho')$ across single, double, and triple Compton scattering scenarios offers a comprehensive picture of how environmental interactions influence quantum superpositions, complementing and extending the entanglement-based analysis.

\subsection{Single Compton Scattering Process}
For a single Compton scattering event on quantum coherence, the $l_1$-norm measure is computed from the evolved density matrix $\rho'$ reconstructed via the Stokes–Mueller formalism described in Sec. 2. The $l_1$-norm of a single Compton scattering process is defined in Appendix \ref{AppendixB} (Eq. \ref{l1norm}).

Figure \ref{l1norm1} shows the dependence of $C_{l_1}(\rho')$ on the polar scattering angle $\theta_{21}$. At zero scattering angle ($\theta_{21}=0^\circ$), the coherence starts from the value $C_{l_1}(\rho')=1$. With increasing $\theta_{21}$, the $l_1$-norm rises and reaches its maximum value ($C_{l_1}(\rho')\approx2.1$, around $\theta_{21}\approx76^\circ$), indicating that polarization superposition is best preserved in this angular region. For larger scattering angles, the coherence gradually decreases and reaches its minimum value ($C_{l_1}(\rho')=0.6$) in the backscattering regime.
A central observation is that coherence remains relatively large ($\approx 2$) near $\theta_{21}\approx90^\circ$, precisely where the concurrence vanishes. 
Thus, single Compton scattering cannot fully eliminate coherence, even though it can completely destroy entanglement (in sharp contrast to entanglement, which disappears near $90^\circ$).

\begin{figure}[h!]
	\centering
	\begin{subfigure}{0.49\textwidth}
		\centering
		\includegraphics[width=\textwidth]{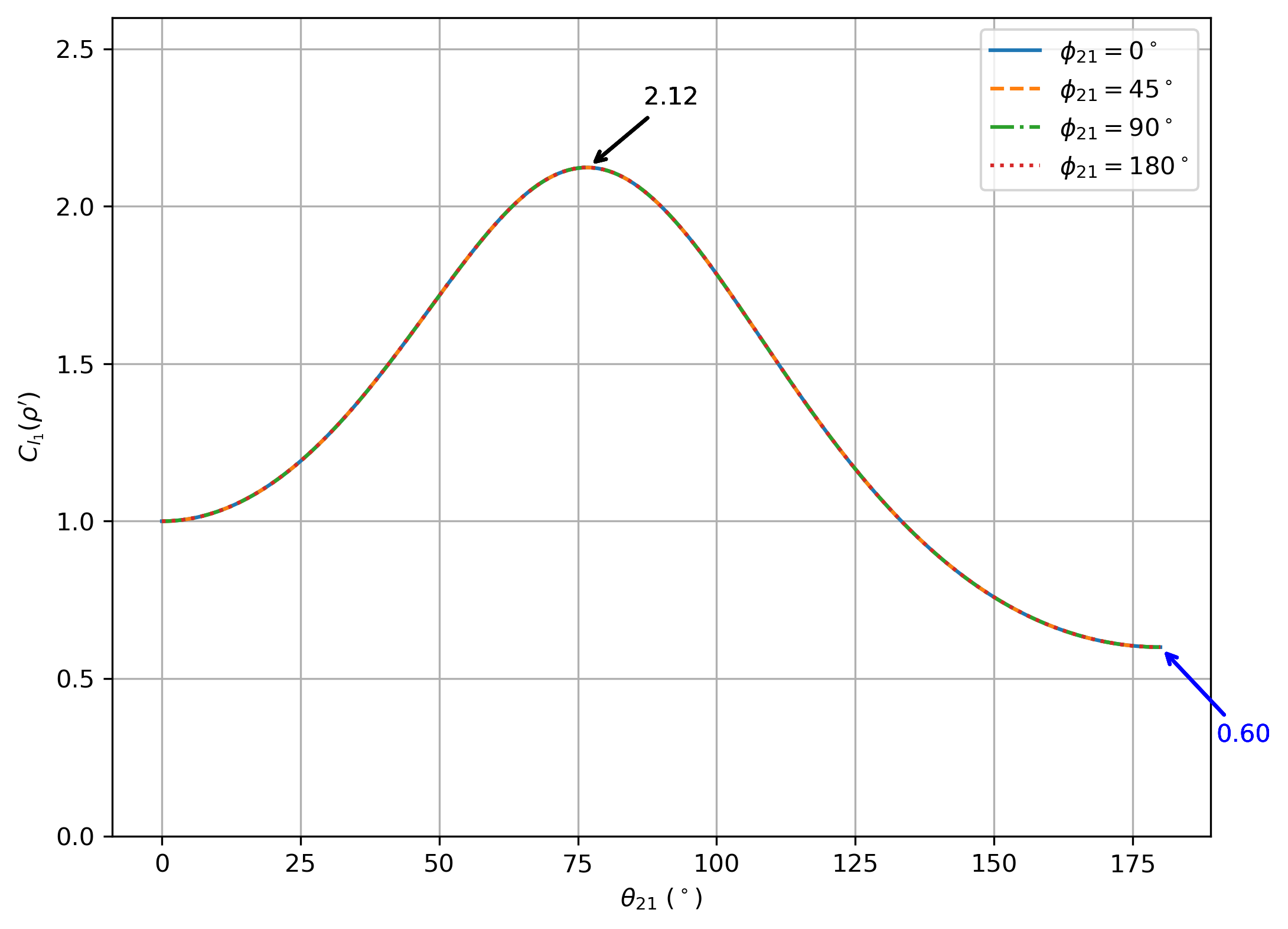}
	\end{subfigure}
	\hfill
	\begin{subfigure}{0.49\textwidth}
		\centering
		\includegraphics[width=\textwidth]{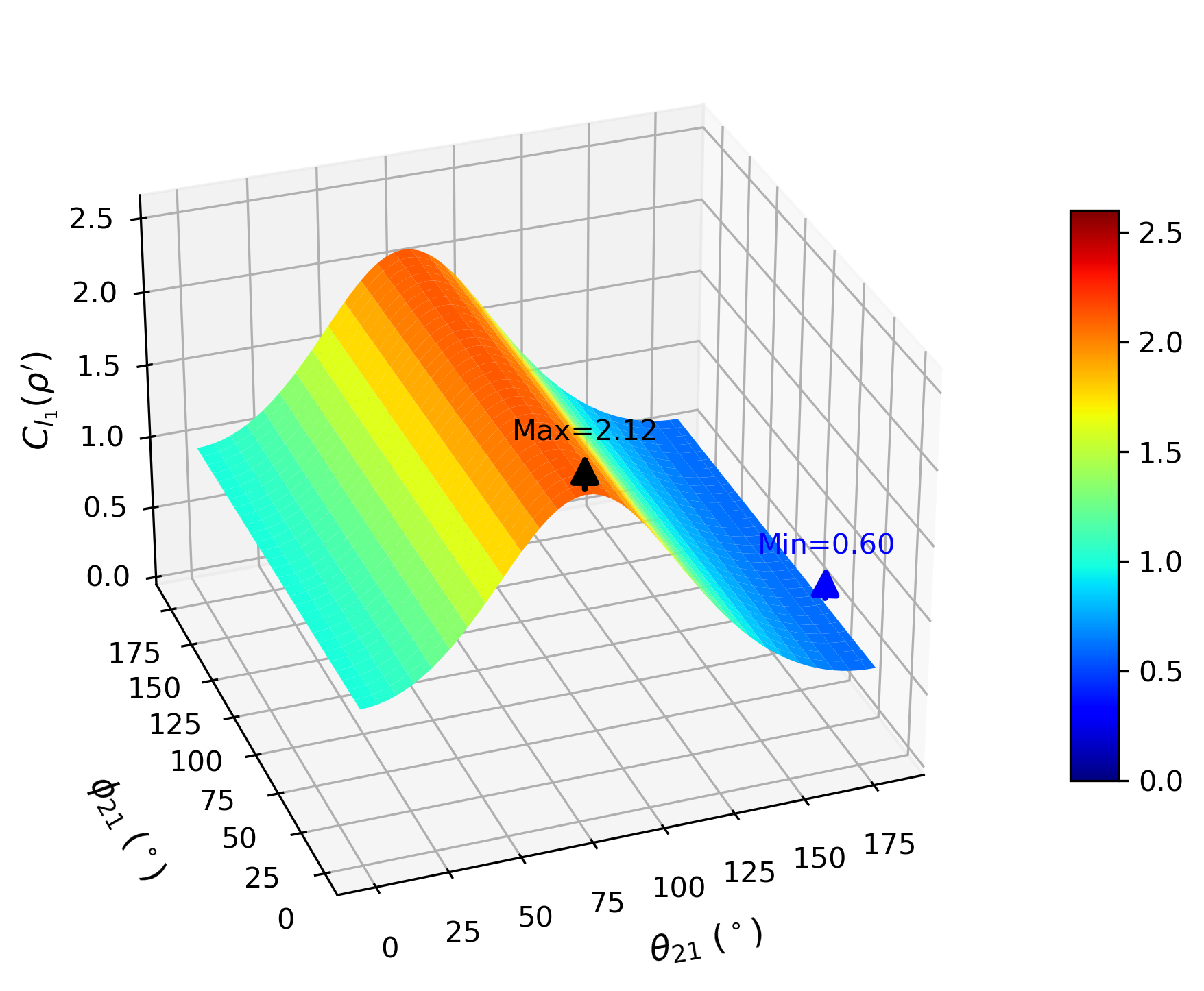}
	\end{subfigure}
	\caption{\footnotesize  Quantum coherence $C_{l_1}(\rho')$ for single Compton scattering as a function of the polar angle $\theta_{21}$ (left), and $\theta_{21}$ and $\phi_{21}$ (right). The coherence peaks near $75^\circ$ and decreases towards backscattering angles while remaining independent of $\phi_{21}$. }
	\label{l1norm1}
\end{figure}

\begin{figure}[h!]
	\centering
	\includegraphics[width=0.6\textwidth]{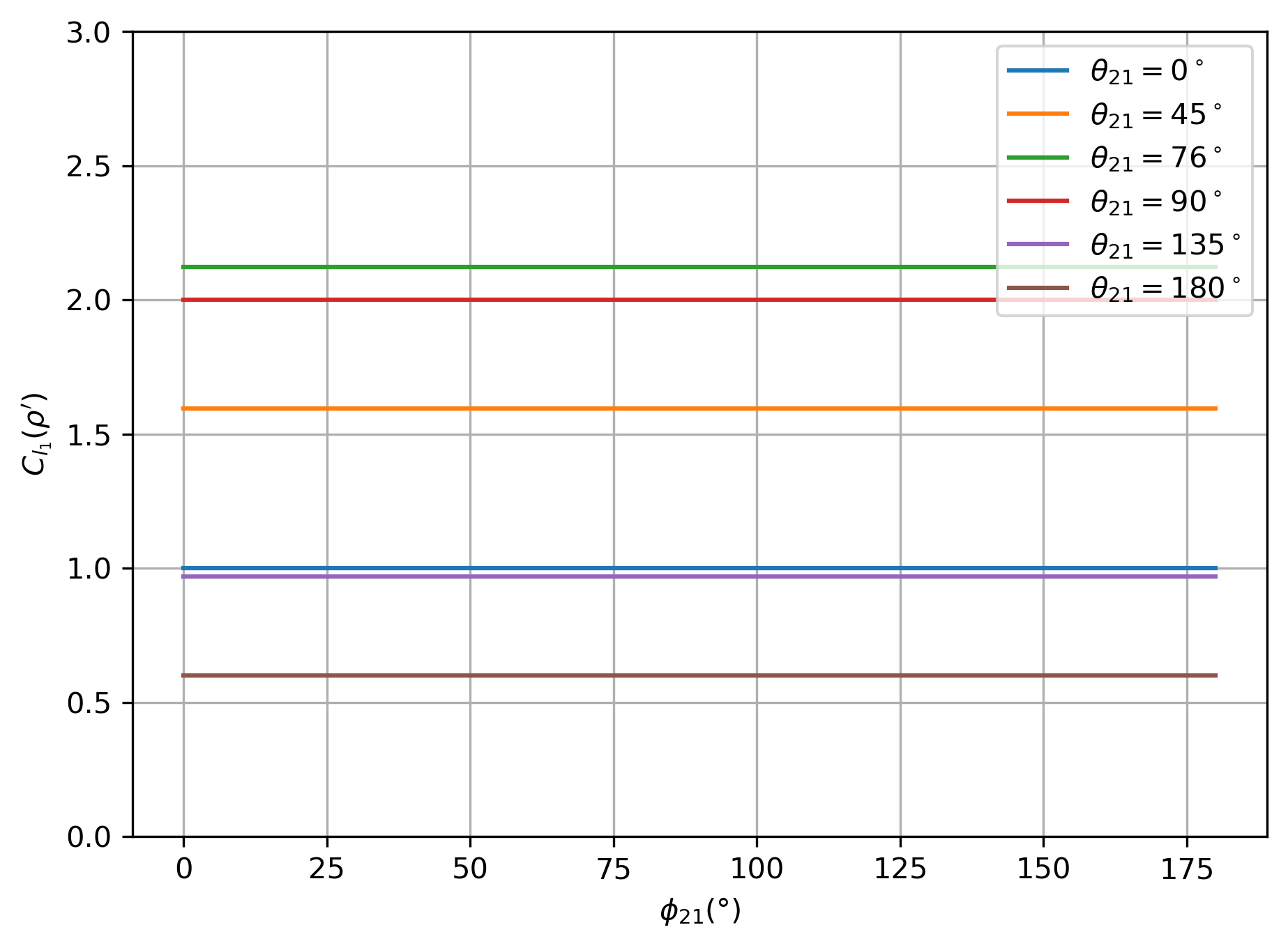}
	\caption{\footnotesize Quantum coherence $C_{l_1}(\rho')$ for single Compton scattering as a function of the azimuthal angle $\phi_{21}$ for several fixed polar angles. The coherence remains constant for all $\phi_{21}$, confirming complete azimuthal invariance.}
	\label{l1norm1phi}
\end{figure}

Moreover, as shown in Fig. \ref{l1norm1phi}, the $l_1$-norm is completely independent of the azimuthal angle $\phi_{21}$, reflecting that azimuthal rotations do not affect coherence under single scattering. This matches the single scattering entanglement behavior and reflects the purely local unitary nature of azimuthal rotations.

\subsection{Double Compton Scattering Process}

\begin{figure}[h!]
	\centering
	\begin{subfigure}{0.49\textwidth}
		\centering
		\includegraphics[width=\textwidth]{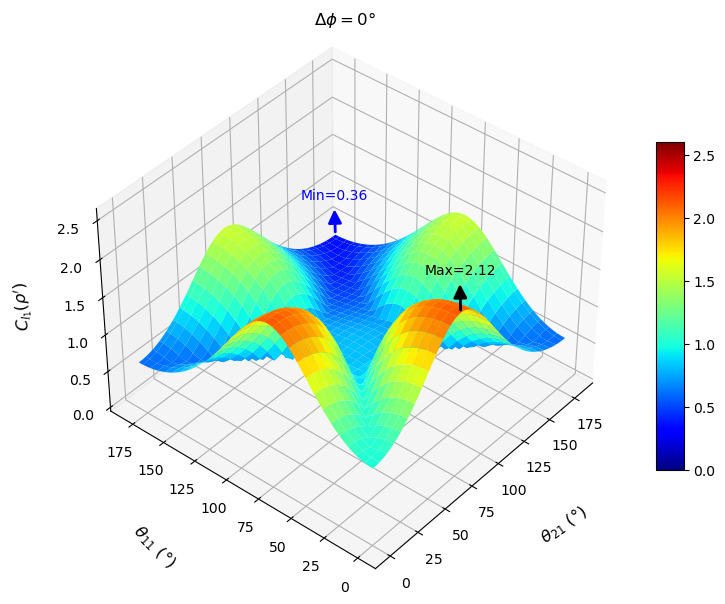}
	\end{subfigure}
	\hfill
	\begin{subfigure}{0.49\textwidth}
		\centering
		\includegraphics[width=\textwidth]{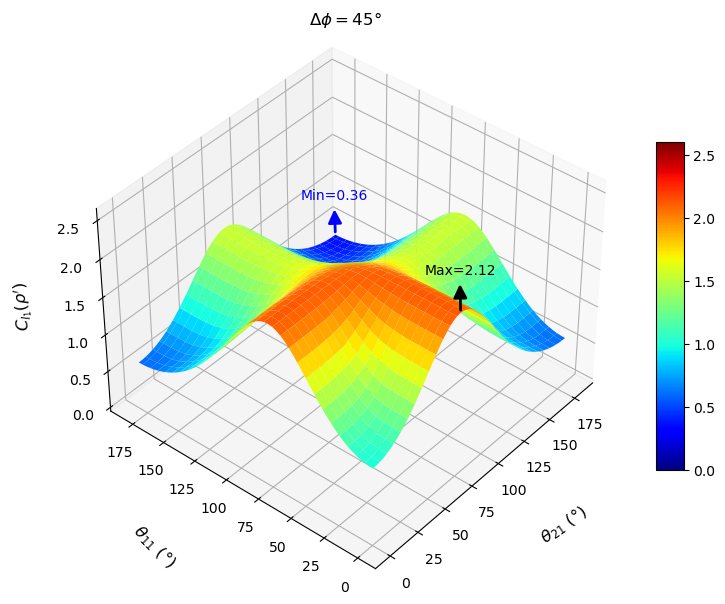}
	\end{subfigure}
	\hfill
	\begin{subfigure}{0.49\textwidth}
		\centering
		\includegraphics[width=\textwidth]{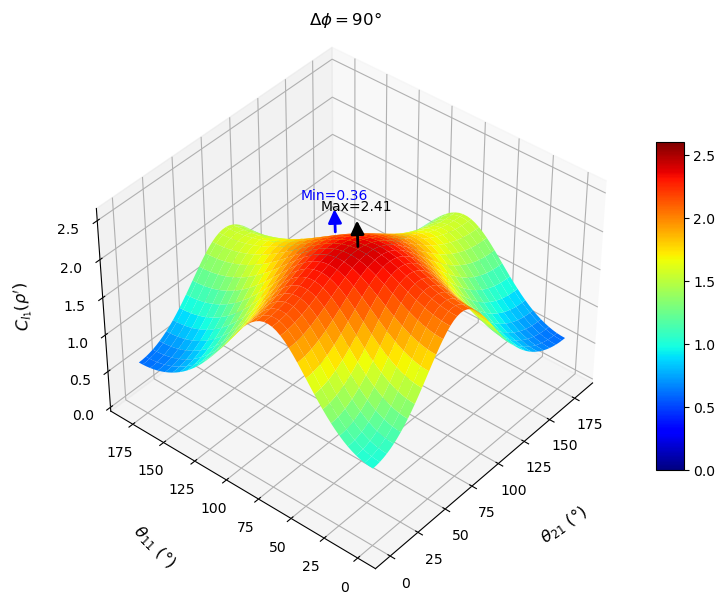}
	\end{subfigure}
	\hfill
	\begin{subfigure}{0.49\textwidth}
		\centering
		\includegraphics[width=\textwidth]{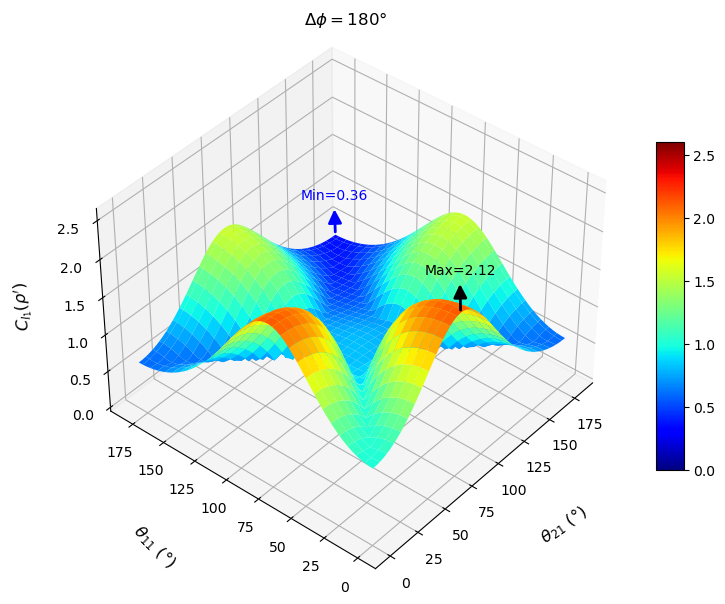}
	\end{subfigure}
	\caption{\footnotesize  Quantum coherence $C_{l_1}(\rho')$ for double Compton scattering as a function of $\theta_{11}$ and $\theta_{21}$ for several azimuthal differences $\Delta\phi$. Coherence peaks near $\approx 75^\circ$ and decreases for large polar angles.}
	\label{l1norm2}
\end{figure}

When each photon scatters once, coherence exhibits a richer dependence on the two polar angles $\theta_{11}$ and $\theta_{21}$ and the azimuthal angle difference $\Delta\phi$. Figure \ref{l1norm2} shows the coherence surface for several values of $\Delta\phi$.

Coherence starts from $C_{l_1}(\rho')$ at small angles and reaches its maximum when either polar angle lies near $ 67^\circ$. For large angles, especially when both photons approach the backscattering limit, coherence decreases and approaches its minimum.

\begin{figure}[h!]
	\centering
	\includegraphics[width=0.6\textwidth]{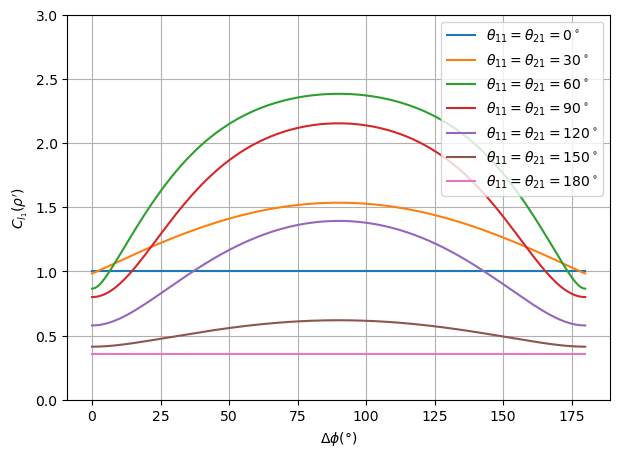}
	\caption{\footnotesize Quantum coherence $C_{l_1}(\rho')$ for double Compton scattering as a function of $\phi_{21}$ for several fixed polar angles. Azimuthal angle dependence appears for $\theta\gtrsim30^\circ$ and disappears at $\theta=180^\circ$.}
	\label{l1norm2phi}
\end{figure}

In contrast to the single scattering case, the $l_1$-norm now depends significantly on $\Delta\phi$, as seen in Fig. \ref{l1norm2phi}. The strongest dependence occurs for intermediate angles ($\theta\gtrsim30^\circ$), while the coherence becomes independent at $\theta=0^\circ$ and $180^\circ$. Minimum coherence typically appears near $\Delta\phi= 180^\circ$, corresponding to anti-parallel scattering geometries where decoherence is enhanced. 

\subsection{Triple Compton Scattering Process}

\begin{figure}[h!]
 	\centering
 	\begin{subfigure}{0.49\textwidth}
 		\centering
 		\includegraphics[width=\textwidth]{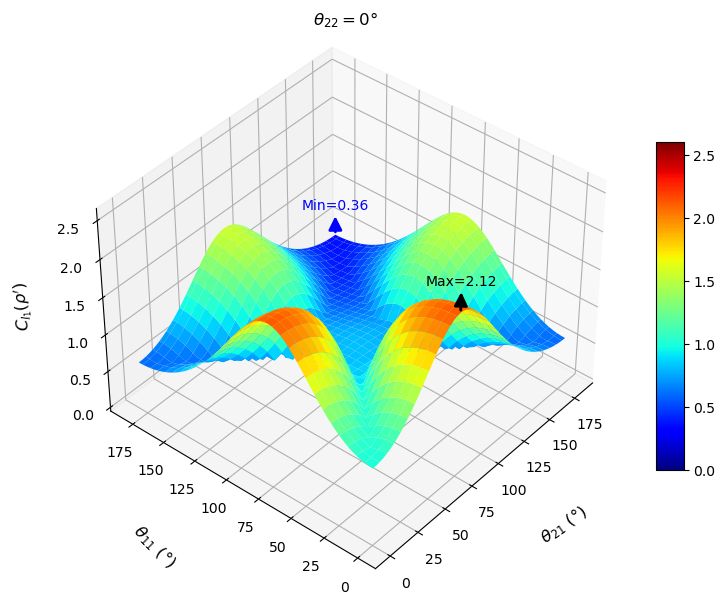}
 	\end{subfigure}
 	\hfill
 	\begin{subfigure}{0.49\textwidth}
 		\centering
 		\includegraphics[width=\textwidth]{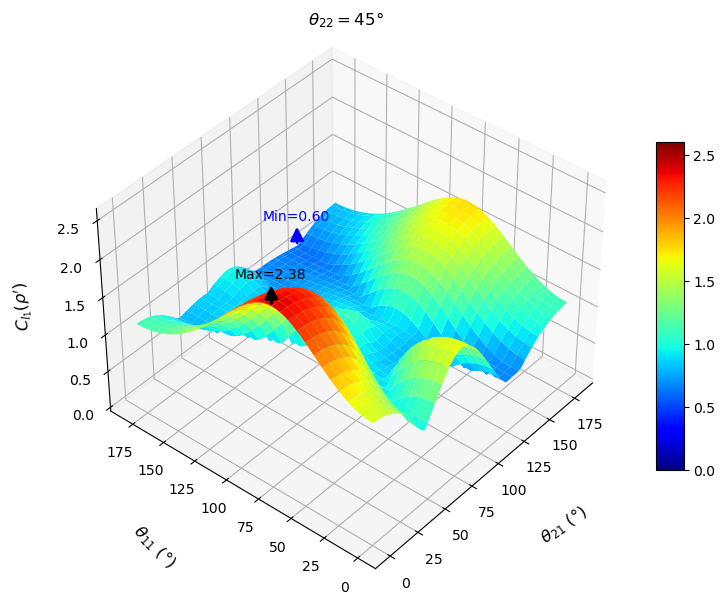}
 	\end{subfigure}
 	\hfill
 	\begin{subfigure}{0.49\textwidth}
 		\centering
 		\includegraphics[width=\textwidth]{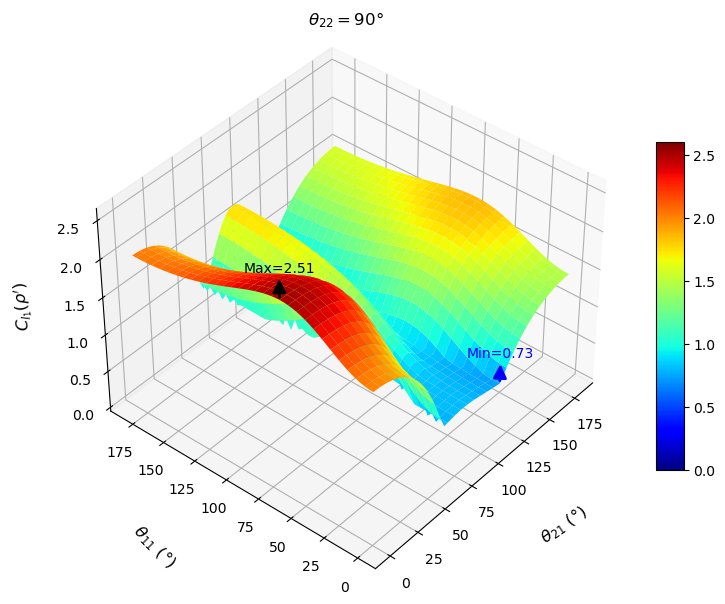}
 	\end{subfigure}
 	\hfill
 	\begin{subfigure}{0.49\textwidth}
 		\centering
 		\includegraphics[width=\textwidth]{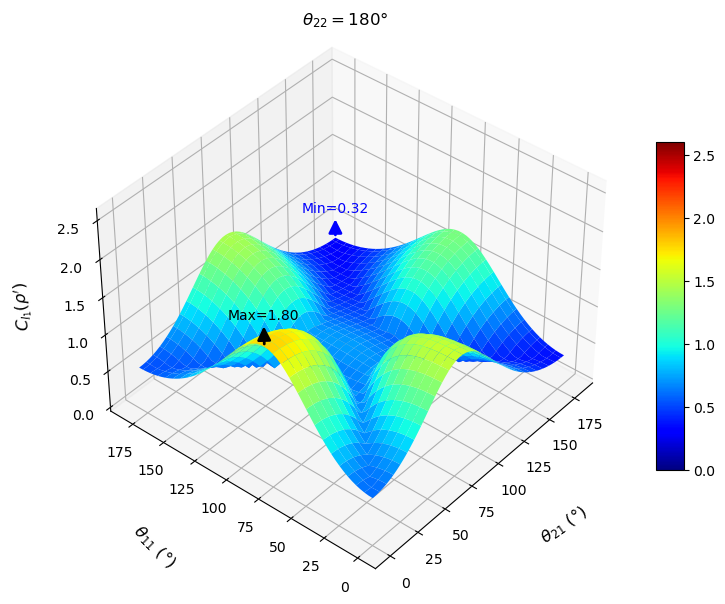}
 	\end{subfigure}
 	\caption{\footnotesize  Quantum coherence $C_{l_1}(\rho')$ for triple Compton scattering as a function of $\theta_{11}$ and $\theta_{21}$ for several $\theta_{22}$. The coherence peak shifts with $75^\circ$ and decreases for large values of $\theta_{22}$.}
 	\label{l1norm3}
 \end{figure}

In the triple Compton scattering scenario, photon 1 scatters once and photon 2 scatters twice. The coherence depends on three polar angles ($\theta_{11}$, $\theta_{21}$, and $\theta_{22}$) and two azimuthal angles ($\Delta\phi$ and $\phi_{22}$). For clarity, we consider $\Delta\phi = \phi_{22}=0$, and $C_{l_1}$ is illustrated as a function of $\theta_{11}$ and $\theta_{21}$ for several values of $\theta_{22}$ (see Fig. \ref{l1norm3}). 

For $\theta_{22}=0^\circ$, the behavior resembles the double Compton scattering case. As $\theta_{22}$ increases, the coherence peak shifts toward $\theta_{11}\approx70^\circ$ while staying at $\theta_{21}\approx0^\circ$. The maximum coherence occurs for $\theta_{22}=90^\circ$, and decreases systematically for $\theta_{22}>90^\circ$, reflecting stronger decoherence in the backscattering regime.

Remarkably, the $l_{1}$-norm never vanishes, even after three scattering events and even in configurations where concurrence collapses completely. The minimum value observed is $0.32$, demonstrating substantial resilience of coherence.

\begin{figure}[h!]
	\centering
	\includegraphics[width=0.6\textwidth]{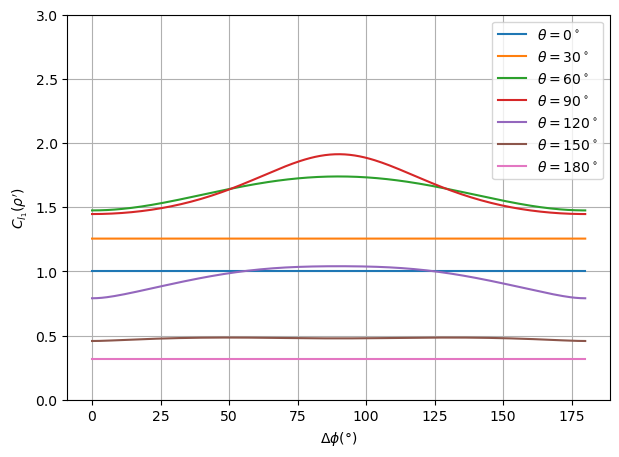}
	\caption{\footnotesize  Quantum coherence $C_{l_1}(\rho')$ for triple Compton scattering as a function of $\Delta\phi$ for several values of polar angles. Nonlinear azimuthal dependence appears for $\theta\gtrsim30^\circ$, but disappears at $\theta=180^\circ$.}
	\label{l1norm3phi}
\end{figure}

The dependence on the azimuthal angle difference $\Delta\phi$ persists in the triple scattering process (Fig. \ref{l1norm3phi}). The $C_{l_1}(\rho')$ exhibits a nonlinear variation with $\Delta\phi$, reaching its maximum at $\theta\approx90^\circ$. In contrast, $\theta=180^\circ$ leads to the minimum coherence values due to the enhancement of decoherence effects in antiparallel scattering configurations. This confirms that the relative scattering geometry plays a crucial role in determining the survivability of quantum coherence in multi scattering regimes.

\section*{Discussion}

The interplay between quantum entanglement and coherence in a photon pair from p-Ps annihilation can be analyzed using quantum resource theories \cite{Streltsov2015,Chitambar2016,Zhu2017}. 
During Compton scattering, part of the photon's polarization information is transferred to the medium electrons. This process couples the photons to an environment, including non-unitary evolution. As a result, nonlocal correlations are degraded, leading to partial or complete loss of entanglement, while local polarization superpositions can persist, yielding finite quantum coherence.

For single scattering events, the concurrence decreases with increasing polar scattering angle, vanishing near $\theta\approx90^\circ$ (Fig. \ref{concurrence13d}). In contrast, the $l_1$-norm of coherence remains relatively large (Fig. \ref{l1norm1}). This demonstrates the distinct sensitivities of entanglement and coherence to polarization-dependent scattering.
For double and triple scattering, the central position of the zero-concurrence region remains fixed near right-angle scattering, indicating that geometry primarily governs the onset of complete disentanglement. Successive scattering events mainly broaden this region and reduce the concurrence magnitude, reflecting its nonlocal nature, which is more delicate to information leakage to the environment than local coherence.

Importantly, coherence can survive when concurrence vanishes. This should be interpreted as the persistence of local polarization coherence rather than residual two-photon entanglement. This observation is consistent with operational results demonstrating that while global coherence is required for entanglement, local coherence can remain robust under decoherence processes. The complementarity observed highlights coherence as a more robust quantum resource under polarization-dependent environmental interactions. 

The degradation of entanglement also depends on the medium. 
Since the probability and angular distribution of Compton scattering are influenced by material electron density and scattering cross sections, concurrence is indirectly affected by the medium through which photons propagate. Regions with higher scattering tendencies cause more frequent or larger-angle interactions, leading to stronger entanglement reduction. In this sense, concurrence can serve as a quantitative indicator of environmental perturbations on the entangled photon pair.

From an application perspective, these results are relevant for quantum-enhanced positron emission tomography (PET). While entanglement may be fragile in scattering-dominated 
media, the persistence of polarization coherence suggests that useful quantum features can still be exploited. Analyzing both coherence and entanglement provides a more complete understanding of environmental effects and can guide detector and imaging materials optimization.

The theoretical framework used here is closely aligned with recent advancements in experimental quantum optics and medical imaging. While earlier studies suggested that Compton scattering in tissue rapidly destroys polarization entanglement \cite{Mcnamara, Moskal2018}, recent experimental evidence shows that entanglement can persist after scattering for angles up to $50^\circ$ \cite{Ivashkin2023, Parashari2024}, with significant disentangling occurring only at larger angles \cite{Bordes2024}. These findings provide empirical support for our model, which quantifies how quantum correlations evolve under sequential scattering.
	
Furthermore, the J-PET scanner enables reconstruction of the scattering plane for each photon using plastic scintillators with excellent timing resolution \cite{Moskal2016, Raj2018}. This capability allows for event-by-event reconstruction of polarization information, making it possible to experimentally probe entanglement in realistic geometries.
	
The persistence of both entanglement and quantum coherence after scattering, as shown in our results, suggests that quantum information is not completely erased by interactions within the medium. In the emerging "Quantum PET" paradigm, the quantum correlation measures considered here could serve as sensitive probes of the molecular environment, such as oxygen concentration or metabolic activity,  consistent with developments in Positronium imaging \cite{Moskal2021}. Thus, our model provides the necessary theoretical foundation for interpreting such quantum signatures in next-generation PET systems.
	
Overall, this study clarifies the distinct roles of entanglement and coherence in polarization-entangled annihilation photons. Coherence emerges as a robust quantum resource even when entanglement is completely lost, providing insights for practical imaging and detection scenarios involving multiple scattering processes.

\section*{Conclusion}
We have analyzed the evolution of polarization entanglement and quantum coherence in annihilation photon pairs undergoing single and multiple Compton scattering events. By explicitly constructing and evaluating the two-photon density matrix for up to three sequential scattering interactions, this work extends previous analyses that were limited to the single scattering regime and provides a systematic framework for tracking quantum state degradation under realistic scattering geometries. Moreover, the quantitative comparison of entanglement and local polarization coherence across sequential scattering events provides insight that has not been explored in previous PET related theoretical studies.

Our results demonstrate that entanglement is highly sensitive to scattering geometry, with the polar angles playing a dominant role. The angular regions associated with complete disentanglement remain centered around right-angle scattering, while additional scattering events mainly enhance entanglement degradation and slightly broaden the zero-concurrence interval, highlighting the strong geometric sensitivity of this nonlocal quantum resource.

In contrast, quantum coherence exhibits remarkable robustness. Even when entanglement is completely destroyed, the $l_1$-norm remains finite, indicating that local polarization superpositions can survive multiple scattering interactions. This behavior is consistent with resource-theoretic results showing that local coherence can survive even when nonlocal entanglement is completely suppressed \cite{Streltsov2015, Chitambar2016}. 

These findings have direct relevance for quantum-enhanced PET imaging. While entanglement may be strongly degraded in realistic, scattering-dominated media, polarization coherence can maintain useful quantum features, such as improved timing or scattered photon discrimination. By analyzing both resources simultaneously, one can better assess the viability of quantum effects in practical imaging scenarios.

In summary, this work highlights the distinct and complementary roles of entanglement and coherence in the annihilation photon pair and provides a framework for evaluating quantum resources under geometry-dependent scattering processes in realistic media.

\appendix

\section{Mueller and rotation matrices for Compton scattering}\label{AppendixA}

In this appendix, we summarize the polarization-dependent Compton scattering formalism using Mueller and rotation matrices. While these relations are standard \cite{Fano, Mcmaster}, here we employ them to construct the Stokes matrix of scattered photons, which serves as the foundation for building the two-photon density matrix. This approach allows us to quantify the evolution of polarization correlations and entanglement in the annihilation photon pair under realistic scattering scenarios.

The Mueller matrix $T_{pj}(\theta)$ describing Compton scattering of a photon from an unpolarized electron target is given by \cite{Fano,Mcmaster}
\begin{equation}
	T_{pj} = \frac{r_0^2}{2} \left( \frac{E_{pj}}{E_{p(j-1)}} \right)^2
	\begin{pmatrix}
		t_{11}^{(pj)} & t_{12}^{(pj)} & 0 & 0 \\
		t_{12}^{(pj)} & 2-t_{12}^{(pj)} & 0 & 0 \\
		0 & 0 & t_{33}^{(pj)} & 0 \\
		0 & 0 & 0 & t_{44}^{(pj)}
	\end{pmatrix},
\end{equation}
where $1\leq j\leq \eta$, with $\eta$ the total number of Compton scattering events. The index $p=1, 2$ labels the photon within the two-photon system, and $r_0$ denotes the classical electron radius. 

The energies $E_{pj}$ and $E_{p(j-1)}$ correspond to the photon energy after and before the $j$-th scattering event, respectively, and are related by the Compton formula:
\begin{equation}\label{16e}
	E_{pj}(\theta)=\frac{E_{p(j-1)}}{1+E_{p(j-1)}(1-\cos\theta_{pj})}\,\,\,\,(1< j\leq\eta),
\end{equation}
with the initial photon energy chosen as $E_0=1$ (corresponding to $511$ keV).

The elements of the transition matrix are defined as:
\begin{equation*}
	\begin{aligned}
		t_{11}^{(pj)}& = 1 + \cos^2\theta_{pj} + \left(E_{p(j-1)} - E_{pj} \right)(1 - \cos\theta_{pj}),\\
		t_{12}^{(pj)}& = t_{21}^{(pj)} = \sin^2\theta_{pj},\\
		t_{22}^{(pj)} &= 2 - t_{12}^{(pj)} = 2 - \sin^2\theta_{pj},\\
		t_{33}^{(pj)} &= 2\cos\theta_{pj},\\
		t_{44}^{(pj)} &= 2 \cos\theta_{pj} + \left(E_{p(j-1)} - E_{pj} \right)(1 - \cos\theta_{pj})\cos\theta_{pj}.
	\end{aligned}
\end{equation*}
Where $0\leq\theta_{pj}\leq\pi$ is the scattering angle for the $j$-th interaction.

Since Stokes parameters depend on the choice of polarization reference frame, an azimuthal rotation of the coordinate axes by an angle $\phi_{pj}$ is represented by the rotation matrix \cite{Mcmaster}
\begin{equation}
	M_{pj} = 
	\begin{pmatrix}
		1 & 0 & 0 & 0 \\
		0 & \cos2\phi_{pj} &  \sin2\phi_{pj} & 0 \\
		0 & - \sin2\phi_{pj} &  \cos2\phi_{pj} & 0 \\
		0 & 0 & 0 &1
	\end{pmatrix},
\end{equation}
with $0\leq\phi_{pj}\leq 2\pi$. These rotation matrices ensure that the polarization vector is expressed in the correct local coordinate frame for each successive Compton interaction.

\section{Single scattering density matrix, concurrence, and $l_{1}$-norm}\label{AppendixB}
Here, we explicitly construct the density matrix for the single Compton scattering case using the Mueller and rotation matrices from Appendix A. The resulting matrix $\rho'$ captures the polarization state of the scattered photon and, together with the unscattered partner, provides a complete description of the two-photon system. This enables direct evaluation of concurrence and the $l_1$-norm. Substituting the corresponding Mueller and rotation matrices into Eq. (\ref{density matrix}) yields
\begin{equation}\label{dm}
	\rho'=\frac{1}{4t_{11}^{(21)}}\scriptsize
	\left[
	\setlength{\arraycolsep}{2pt}
	\begin{matrix}
		1 -t_{44}^{(21)} &  t_{12}^{(21)}& - t_{12}^{(21)} e^{- 2 i \phi_{21}} &  { (t_{33}^{(21)} + t_{12}^{(21)} - 2) e^{- 2 i \phi_{21}}}\\
		t_{12}^{(21)} & 1 +  t_{44}^{(21)} & (-t_{33}^{(21)} + t_{12}^{(21)} - 2) e^{- 2 i \phi_{21}} & - t_{12}^{(21)} e^{- 2 i \phi_{21}}\\[2mm]
		-t_{12}^{(21)} e^{2 i \phi_{21}} &  (-t_{33}^{(21)}  +  t_{12}^{(21)} - 2) e^{2 i \phi_{21}}& 1 +  t_{44}^{(21)} &  t_{12}^{(21)}\\
		(t_{33}^{(21)}+t_{12}^{(21)} - 2) e^{2 i \phi_{21}} & - t_{12}^{(21)} e^{2 i \phi_{21}} &  t_{12}^{(21)} & 1 -  t_{44}^{(21)}
	\end{matrix}
	\right],
\end{equation}
where $\phi_{21}$ is a real azimuthal rotation angle associated with the scattering geometry of photon 2.

Using the standard Wootters formula, the concurrence of $\rho'$  for the single scattering scenario is found to be	
\begin{equation}\label{con}
	\begin{aligned}
			C(\rho') &=
			\sqrt{\dfrac{1}{16\,(t_{11}^{(21)})^2}\,\bigl(A + 2\sqrt{B}\bigr)}
			\;-\;
			\sqrt{\dfrac{1}{16\,(t_{11}^{(21)})^2}\,\bigl(A - 2\sqrt{B}\bigr)}
			\\[2mm]
			&\quad
			- \left|
			\frac{-2 + t_{11}^{(21)} + t_{12}^{(21)} + t_{33}^{(21)} - t_{44}^{(21)}}
			{4\,t_{11}^{(21)}}
			\right|
			- \left|
			\frac{-2 + t_{11}^{(21)} + t_{12}^{(21)} - t_{33}^{(21)} + t_{44}^{(21)}}
			{4\,t_{11}^{(21)}}
			\right|,
	\end{aligned}
\end{equation}
where
\begin{align}\nonumber
	A &= 4 + (t_{11}^{(21)})^2
	- 2\,t_{11}^{(21)}(-2 + t_{12}^{(21)})
	- t_{12}^{(21)}(4 + 3\,t_{12}^{(21)})
	+ (t_{33}^{(21)}+t_{44}^{(21)})^2,\\[2mm]
	B &= (2 + t_{11}^{(21)} - 3\,t_{12}^{(21)})
	(2 + t_{11}^{(21)} + t_{12}^{(21)})
	(t_{33}^{(21)}+t_{44}^{(21)})^2.
\end{align}
			
The $l_{1}$-norm of coherence is defined as the sum of the absolute values of all off-diagonal elements of $\rho'$. For the single scattering case, this reduces to
\begin{equation}\label{l1norm}
	C_{l_1}(\rho')= {\frac{1}{2|t_{11}^{(21)}|}} (4|{t_{12}^{(21)}} |+ |2-t_{12}^{(21)}+t_{33}^{(21)}| +|-2+t_{12}^{(21)}+t_{33}^{(21)}|).
\end{equation}

\end{document}